\documentclass[]{aa} 
\usepackage{graphicx}
\usepackage[version=4]{mhchem}
\usepackage{txfonts}
\usepackage{amsmath}
\usepackage{pifont}
\usepackage{rotating}
\usepackage{textcomp}
\usepackage{natbib}
\usepackage{pdflscape}
\begin{document}

   \title{Formation of interstellar methanol ice prior to the heavy \ce{CO} freeze--out stage}
   \author{D. Qasim\inst{1},
   K.--J. Chuang\inst{1},
   G. Fedoseev\inst{2},
   S. Ioppolo\inst{3,4},
   A.C.A. Boogert\inst{5},
          \and
          H. Linnartz\inst{1}
          }

   \institute{Sackler Laboratory for Astrophysics, Leiden Observatory, Leiden University, PO Box 9513, NL--2300 RA Leiden, The Netherlands\\
              \email{dqasim@strw.leidenuniv.nl}
             \and INAF--Osservatorio Astrofisico di Catania, via Santa Sofia 78, 95123 Catania, Italy \
             \and School of Electronic Engineering and Computer Science, Queen Mary University of London, Mile End Road, London E1 4NS, UK
             \and School of Physical Sciences, STEM, Open University, Milton Keynes MK7 6AA, UK
             \and Institute for Astronomy, University of Hawaii at Manoa, 2680 Woodlawn Drive, Honolulu, HI 96822--1839 
             }
             
\date{Received X; accepted Y}

 
  \abstract
   {The formation of methanol (\ce{CH3OH}) on icy grain mantles during the star formation cycle is mainly associated with the \ce{CO} freeze--out stage. Yet there are reasons to believe that \ce{CH3OH} also can form at an earlier period of interstellar ice evolution in \ce{CO}--poor and \ce{H2O}--rich ices.}
   {This work focuses on \ce{CH3OH} formation in a \ce{H2O}--rich interstellar ice environment following the \ce{OH}--mediated \ce{H}--abstraction in the reaction, \ce{CH4 + OH}. Experimental conditions are systematically varied to constrain the \ce{CH3OH} formation yield at astronomically relevant temperatures.}
   {\ce{CH4}, \ce{O2}, and hydrogen atoms are co--deposited in an ultrahigh vacuum chamber at 10--20 K. \ce{OH} radicals are generated by the \ce{H + O2} surface reaction. Temperature programmed desorption -- quadrupole mass spectrometry (TPD--QMS) is used to characterize \ce{CH3OH} formation, and is complemented with reflection absorption infrared spectroscopy (RAIRS) for \ce{CH3OH} characterization and quantitation.}
   {\ce{CH3OH} formation is shown to be possible by the sequential surface reaction chain, \ce{CH4 + OH}$\,\to\,$\ce{CH3 + H2O} and \ce{CH3 + OH}$\,\to\,$\ce{CH3OH} at 10--20 K. This reaction is enhanced by tunneling, as noted in a recent theoretical investigation \citep{lamberts2017importance}. The \ce{CH3OH} formation yield via the \ce{CH4 + OH} route versus the \ce{CO + H} route is approximately 20 times smaller for the laboratory settings studied. The astronomical relevance of the new formation channel investigated here is discussed.}
   {}
   

\authorrunning{Qasim et al.}
  
   \maketitle
%

\section{Introduction}

Methanol (\ce{CH3OH}) is an important interstellar molecule. \ce{CH3OH} has been observed abundantly in both the gas phase \citep{friberg1988methanol,turner1998physics,parise2002detection,bergman2011deuterated,wirstrom2011observational,guzman2013iram,oberg2014complex,taquet2015constraining} and the solid state \citep{grim1991detection,allamandola1992infrared,skinner1992methanol,chiar1996three, dartois1999methanol,gibb2000inventory,pontoppidan2003detection,taban2003stringent,gibb2004interstellar,boogert2008c2d,bottinelli2010c2d,boogert2011ice}. It is generally accepted that \ce{CH3OH} formation is most efficient by solid state interactions on icy grain mantles. Models, supported by experiments, of gas phase synthesis of \ce{CH3OH} provide abundances orders of magnitude below the observed fractional abundance of \ce{CH3OH} \citep{garrod2006gas,geppert2006dissociative}, while solid state laboratory studies show that \ce{CH3OH} is efficiently formed in \ce{CO}--rich ices through sequential hydrogenation of \ce{CO} \citep{hiraoka1994formation,watanabe2002efficient,fuchs2009hydrogenation}. This is further supported by computational models \citep{cuppen2009microscopic,fuchs2009hydrogenation,garrod2006gas} that show that hydrogenation of \ce{CO} ice leads to the production of \ce{CH3OH}. Indeed, these findings are in line with the spectroscopic interpretation of observational data; \ce{CO} and \ce{CH3OH} are found to coexist in \ce{CO}--rich and \ce{H2O}--poor interstellar ices \citep{cuppen2011co,boogert2015observations,penteado2015spectroscopic}. Not only is it a prevalent interstellar molecule, but \ce{CH3OH} is also an important precursor in the formation of larger species. \citet{oberg2009formation} illustrated that upon vacuum UV irradiation, \ce{CH3OH} can break apart into fragments that can recombine to form complex organic molecules (COMs). In UV--rich environments in the interstellar medium, the formation of \ce{CH3OH} may thus be crucial to the formation of larger COMs. The studies by \citet{chuang2015h,chuang2017production} and \citet{fedoseev2017formation} show that \ce{CH3OH} may be a key player in the formation of larger COMs, also for cold dense prestellar core conditions. The authors show that radicals derived from \ce{CH3OH} via an abstraction process can recombine or combine with other radicals, resulting in the formation of COMs as large as glycerol, even at temperatures below 20 K and without the need for UV--induced radiation.

Recent theoretical and experimental efforts \citep{lamberts2017importance} have provided results that have sparked the idea of \ce{H2O} and \ce{CH3OH} coexisting in a \ce{CO}--poor and \ce{H2O}--rich interstellar ice. Such ices are thought to form before the ``heavy'' CO freeze--out stage, which starts to occur at a cloud extinction ($A_V$) of $A_V$ > 3 and dust temperatures < 20 K, and that is followed by the ``catastrophic'' CO freeze--out stage at $A_V$ > 9 and dust temperatures also < 20 K. Prior to the heavy CO freeze--out stage, a \ce{CO}:\ce{H2O} ice ratio of < 5\% is expected due to some CO freezing out as well as atom--addition reactions (C, O, H, etc.). Due to the relatively low gas densities at this stage, only a relatively low amount of CO is able to accrete onto dust grains, as examined in \citet{pontoppidan2006spatial} and discussed in \citet{Oberg2011spitzer}, \citet{boogert2015observations}, and \citet{van2017astrochemistry}. This time period is known as the ``\ce{H2O}--rich ice phase'' or ``polar ice phase'' (i.e., the phase that lacks CO ice). \ce{CH3OH} formation through this reaction, \ce{CH4 + OH}, can take place in this phase. However, this does not exclude this reaction to also take place during the \ce{CO} freeze--out stage, where \ce{CH3OH} formation is normally dominated by \ce{CO} hydrogenation. During this time, \ce{O2} might be intimately mixed with \ce{CO} \citep{vandenbussche1999constraints} and can be hydrogenated to form \ce{OH} radicals. Gas phase \ce{CH4} becomes more dominant relative to atomic carbon (van Dishoeck 1998) and could freeze out with \ce{CO}. The importance of the present study is that it focuses on a phase in which \ce{CH3OH} is only formed through \ce{CH4 + OH}, as \ce{CO} is not frozen out yet, and an identification of \ce{CH3OH} in corresponding environments, therefore, directly relates to this alternative reaction pathway. 

The study by \citet{lamberts2017importance} was done to provide reaction rates at cryogenic temperatures of the reaction between methane (\ce{CH4}) and \ce{OH} as \ce{CH4} has been observed in the \ce{H2O}--rich ice phase \citep{boogert1997infrared,oberg2008c2d}, and \ce{OH} radicals are expected to be abundant in that phase \citep{cuppen2007simulation,ioppolo2008laboratory,cuppen2010water,oba2012water,lamberts2013water}. These authors, as well as others \citep{wada2006methanol,hodyss2009photochemistry,weber2009hydrogen,zins2012reactivity,bossa2015methane}, showed that \ce{CH4} hydrogen abstraction by an \ce{OH} radical results in the efficient formation of \ce{CH3} radicals; a process that can be induced by tunneling \citep{lamberts2017importance}. Thus, these findings can be taken one step further by postulation that the \ce{OH}--induced abstraction reaction of \ce{CH4} may result in the formation of \ce{CH3OH} by the sequential reaction chains: \ce{CH4 + OH}$\,\to\,$\ce{CH3 + H2O} and \ce{CH3 + OH}$\,\to\,$\ce{CH3OH}. There are already a number of laboratory experiments that exhibit \ce{CH3OH} formation via energetic processing of ice mixtures containing \ce{CH4} and \ce{H2O} \citep{schutte1988evolution,d1996iso,moore1998infrared,wada2006methanol,hodyss2009photochemistry,martin2016study}. In addition, \citet{bergner2017methanol} have explored the formation of \ce{CH3OH} via O--atom insertion in \ce{CH4} molecules upon photodissociation of \ce{O2} in binary mixtures. In our paper, the reaction \ce{CH4} + \ce{OH} leading to the formation of \ce{CH3OH} under relevant cold dense core conditions is discussed. The \ce{OH} radicals are already present at the earliest stages of ice formation via photodissociation of \ce{H2O} molecules and by atomic H-- and O--rich accreting gas (see, e.g., \citealt{boogert2015observations}). Therefore, this formation mechanism could work even under non--energetic conditions, and can be extended to the formation of COMs that contain the hydroxyl functional group (--\ce{OH}). Furthermore, \ce{CH3} radical formation is constrained to abstraction reactions only, whereas in the energetic processing studies, it is unclear whether \ce{CH3} radicals are formed by energetic and/or non--energetic pathways.   

   Data from astronomical observations also provide an incentive to explore \ce{CH3OH} ice formation before the heavy CO freeze--out stage (i.e., in the \ce{H2O}--rich ice phase). In the review published by \citet{boogert2015observations}, Figure 7 displays the observed ice column densities of \ce{CH3OH} and \ce{H2O}, as well as other species, as a function of ($A_V$). Consistent with the current model of interstellar ice evolution, \ce{H2O} is concentrated at a lower $A_V$ with respect to \ce{CH3OH}. However, taking a closer look at the figure reveals that \ce{CH3OH} upper limit values can be found at a lower $A_V$ and at lower column densities than \ce{CH3OH} detections. Such upper limits do not prove that \ce{CH3OH} can be formed at lower $A_V$, but they also do not exclude the idea. A possible explanation for the relatively low $A_V$ and column densities for the \ce{CH3OH} upper limits may be due to other reaction pathways expected to take place before the CO freeze--out phase. It is also possible that at such low $A_V$, the sensitivity of the telescopes used may not have been high enough. In the dawn of the James Webb Space Telescope (JWST) era, it is expected that soon it will be possible to extend on the formation schemes in which solid \ce{CH3OH} is formed or consumed. 
   
   In this paper, the formation of \ce{CH3OH} ice by \ce{CH4 + OH} under relevant cold dense core conditions is discussed. Section~\ref{2.} provides an overview of the experimental conditions and methods. Section~\ref{3.} details how \ce{CH3OH} was identified, formed in the ice, and quantified. Section~\ref{4.} provides insights into how the laboratory results presented here can be used to constrain the ice chemistry in the \ce{H2O}--rich ice phase of cold prestellar core environments, as well as a discussion of Figure 7 from \citet{boogert2015observations}. Section~\ref{5.} summarizes the findings from this paper. 

\section{Experimental setup and methods}
\label{2.}

   The experiments described here are performed with SURFRESIDE$^2$. This is a double atom beam line, ultrahigh vacuum (UHV) setup with a base pressure of $\sim$$10^{-10}$ mbar in the main chamber. An in--depth description of the setup and experimental procedures can be found in \citet{ioppolo2013surfreside2} and \citet{linnartz2015atom}. Ices are grown on a gold--plated copper substrate ($2.5 \times 2.5$ cm$^{2}$) that is attached to the cold--finger of a closed--cycle helium cooled cryostat (ColdEdge CH--204 SF). Temperatures as low as 7 K can be realized. The temperature settings are controlled by a LakeShore 340 temperature controller. The substrate temperature is measured by a silicon diode sensor (DT--670) with an absolute accuracy of 1 K. The substrate temperature is changed by resistive heating of a tungsten filament. Incorporation of a sapphire rod within the cryocooler allows substrate temperatures as high as 450 K.
   
   The processes taking place in the ice can be studied through infrared (IR) spectroscopy and mass spectrometry. Species formed in the ice are probed by their IR signatures with a Fourier transform infrared spectrometer (FTIR; Agilent Cary 640/660) applying the reflection absorption infrared spectroscopy (RAIRS) technique. The FTIR permits a coverage of 6000--700 cm$^{-1}$ with a resolution of 1 cm$^{-1}$. To further constrain species present in the ice, a quadrupole mass spectrometer (QMS; Spectra Microvision Plus LM76) is used to measure the desorption temperature of ice species during a temperature programmed desorption (TPD) run, as well as the mass spectrum of the desorbing species upon electron impact ionization. A commonly used electron impact ionization energy of 70 eV is chosen. All TPD experiments involve a linear ramp rate of 2 K min$^{-1}$.   
   
         
        \begin{sidewaystable}
\caption{Experiments performed in this study. All fluxes, except the H flux, are derived from the Hertz--Knudsen equation.} 
\label{table1} 
\centering 
\begin{tabular}{c c c c c c c c c c c}
\hline\hline 
No. & Experiments & Ratio & T$_{sample}$$^{*}$ & Flux$_{\ce{CH4}}$ & Flux$_{\ce{H}}$ & Flux$_{\ce{O2}}$ & Flux$_{\ce{CH3OH}}$ & Flux$_{\ce{H2O}}$ & Flux$_{\ce{CO}}$ & Time\\
 & & \ce{CH4}:\ce{H}:\ce{O2} & (K) & cm$^{-2}$s$^{-1}$ & cm$^{-2}$s$^{-1}$ & cm$^{-2}$s$^{-1}$ & cm$^{-2}$s$^{-1}$ & cm$^{-2}$s$^{-1}$ & cm$^{-2}$s$^{-1}$ & (s) \\ 
\hline 
1.0 & \ce{CH4 + H + O2} & 1:2:1 & 10 & 3E12 & 6E12 & 4E12 & -- & -- & -- & 43200 \\ 
2.0 & \ce{CH4 + H + O2} & 1:2:1 & 10 & 3E12 & 6E12 & 4E12 & -- & -- & -- & 21600\\
2.1 & $^{13}$\ce{CH4 + H + O2} & 1:2:1 & 10 & 3E12 & 6E12 & 4E12 & -- & -- & -- & 21600\\
2.2 & \ce{CH4 + H} + $^{18}$\ce{O2} & 1:2:1 & 10 & 3E12 & 6E12 & 4E12 & -- & -- & -- & 21600\\
2.3 & $^{13}$\ce{CH4} + \ce{H} + $^{18}$\ce{O2} & 1:2:1 & 10 & 3E12 & 6E12 & 4E12 & -- & -- & -- & 21600\\ 
3.0 & \ce{CH3OH} & -- & 10 & -- & -- & -- & 4E13 & -- & -- & 100\\
4.0 & \ce{CH3OH} & -- & 10 & -- & -- & -- & 1E13 & -- & -- & 1200\\
4.1 & \ce{CH3OH} + \ce{O2} (MWAS)$^\diamondsuit$ & -- & 10 & -- & -- & 4E13 & 1E11 & -- & -- & 9000\\
4.2 & \ce{CH3OH} + \ce{CH4} & -- & 10 & 6E13 & -- & -- & 1E11 & -- & -- & 9000\\
4.3 & \ce{CH3OH + H2O} & -- & 10 & -- & -- & -- & 1E11 & 6E13 & --& 9000\\
4.4 & \ce{CH3OH + H2O} (MWAS)$^\diamondsuit$ & -- & 10 & -- & -- & -- & 1E11 & 6E13 & -- & 9000\\
4.5 & \ce{CH3OH + H2O} (MWAS)$^\diamondsuit$ + \ce{CH4} & -- & 10 & 6E13 & -- & -- & 1E11 & 6E13 & -- & 9000\\
4.6 & \ce{CH3OH + CO} & -- & 15 & -- & -- & -- & 2E10 & -- & 7E10 & 3600\\
5.0 & $^{13}$\ce{CH4} + \ce{H} + $^{18}$\ce{O2} & 1:2:1$^\clubsuit$ & 20 & 3E12$^\clubsuit$ & 6E12 & 4E12 & -- & -- & -- & 21600\\
6.0 & \ce{CH4 + H + O2} & 1:2:1 & 10 & 3E12 & 6E12 & 1E13 & -- & -- & -- & 21600\\

\hline 
\end{tabular}

\tablefoot{
\tablefoottext{*}{Temperature of the sample at which the ices are grown.}
\tablefoottext{$\clubsuit$}{The \ce{CH4} flux is lower than the listed value in order to have the same \ce{CH4}:\ce{H2O} ice ratio as found in exp. 2.0. Thus, a constant \ce{CH4} flux is not carried out in this particular experiment.}
\tablefoottext{$\diamondsuit$}{``(MWAS)'' denotes species that were placed in the MWAS chamber. See also Section~\ref{spectral}.}
}
\end{sidewaystable}
   
   In order to simulate cold dense core conditions, all ices are grown on a substrate surface with a temperature between 10--20 K. \ce{CH4}, H, and \ce{O2} are admitted into the main vacuum chamber following the \emph{co--deposition\/} technique (i.e., different molecular species deposited simultaneously), which reproduces interstellar conditions better than with previously used pre--deposition techniques (i.e., different molecular species deposited sequentially) \citep{linnartz2015atom}. Additionally, this technique allows all deposited species to react with one another, regardless of the ice thickness, which is advantageous when trying to probe trace species. The OH radicals are formed by H--atom addition to \ce{O2} \citep{cuppen2010water} and the \ce{CH3} radicals are formed by \ce{OH}--mediated \ce{H}--abstraction of \ce{CH4} (i.e., radicals are formed in situ). A series of control experiments performed by \citet{lamberts2017importance} showed that the produced \ce{OH} radicals are solely responsible for the H--abstraction of \ce{CH4} (i.e., formation of \ce{CH3} radicals directly through reaction with H--atoms is not efficient, as proved in \citet{lamberts2017importance}). We note that the formation of \ce{OH} radicals from \ce{O2} is not necessarily representative of the formation of \ce{OH} radicals in low $A_V$ (low density) environments, where the \ce{O + H} route dominates, as well as photodissociation of \ce{H2O}. Since the \ce{OH} and \ce{CH3} radicals are formed in the ice, they are expected to be thermalized before further reactions occur. 
   
        To perform H--atom addition reactions, H--atoms are produced by a Hydrogen Atom Beam Source (HABS) \citep{tschersich1998formation,tschersich2000intensity,tschersich2008design}, and the H--atom beam line has an angle of 45$^{\circ}$ to surface normal of the gold--plated sample. Hydrogen molecules (\ce{H2}; Praxair 99.8\%) flow into the HABS chamber via a leak valve and are thermally cracked by a tungsten filament. The gas deposition lines of \ce{O2} (Linde Gas 99.999\%) and \ce{CH4} (Linde Gas 99.995\%) are angled at 135$^{\circ}$ and 68$^{\circ}$, respectively, to the plane of the sample's surface. Gas isotopologues, \ce{^{18}O2} (Campro Scientific 97\%) and \ce{^{13}CH4} (Sigma--Aldrich 99\%), are used as controls to aid in identification of the formed ice products.
   
   Fluxes and column densities are characterized as follows. An absolute D--atom flux was measured by \citet{ioppolo2013surfreside2}, and the \ce{O2} and \ce{CH4} fluxes are calculated using the Hertz--Knudsen equation \citep{Kolasinski2012}. Column densities are determined by the relation between absorbance and column density, as described in \citet{hudgins1993mid}. As discussed previously by \citet{ioppolo2013surfreside2}, \citet{teolis2007infrared}, and \citet{loeffler2006synthesis}, such measurements must be done  with caution. Therefore in this work, only the relative column densities of \ce{CH4}, \ce{H2O}, and \ce{CH3OH} are given (i.e., absolute column densities are not listed). Care is taken to use integrated absorbances that do not deviate from the linear trend of the column density over time. The band strength used to determine the \ce{CH3OH} column density is $7.1 \times 10^{-17}$ cm molecule$^{-1}$ (1030 cm$^{-1}$) and is obtained by performance of a He--Ne laser interference experiment in SURFRESIDE$^2$ (Chuang et al. in prep). The underlying experimental procedure is described in detail in \citet{paardekooper2016novel}. \ce{CH4} (1302 cm$^{-1}$) and \ce{H2O} (1659 cm$^{-1}$) band strengths of $8.0 \times 10^{-18}$ cm molecule$^{-1}$ and $9.0 \times 10^{-18}$ cm molecule$^{-1}$, respectively, are extracted from \citet{bouilloud2015bibliographic}. Since the band strengths from \citet{bouilloud2015bibliographic} are obtained from transmission IR experiments, a proportionality factor between the \ce{CH3OH} band strengths from \citet{bouilloud2015bibliographic} and our laser interference experiment is used as a correction factor. It should be noted that literature transmission band strength values cannot be used with optical depth values obtained from a RAIRS experiment partly due to surface--enhanced dipole coupling that occurs in RAIRS. Setup specific values have to be determined for specified conditions. All experiments discussed in this paper are listed in Table~\ref{table1}. 
   
   \section{Results and discussion}
   \label{3.}
   
   \subsection{Identification and analysis of \ce{CH3OH} formation}
   
   In the next sections, three different ways to confirm the formation of \ce{CH3OH} ice by the reaction, \ce{CH4 + OH}, are presented. 
   
   \subsubsection{TPD}
   \label{3.1.1.}

\begin{figure}
\resizebox{\hsize}{!}{\includegraphics{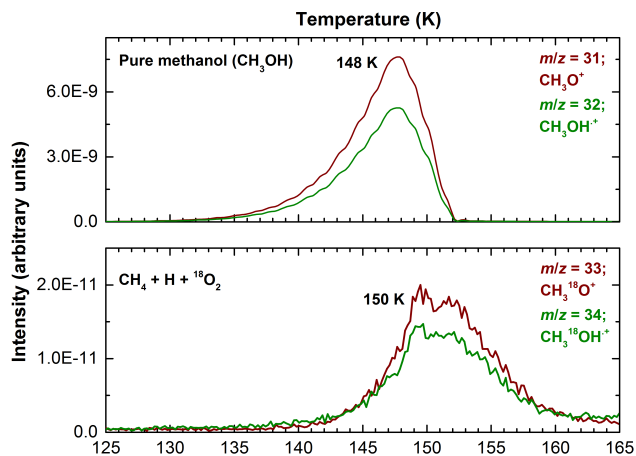}}
\caption{TPD of pure \ce{CH3OH} (top; exp. 3.0) and TPD of the \ce{CH4 + H} + $^{18}$\ce{O2} reaction (bottom; exp. 2.2).} 
\label{fig1}
\end{figure}

Figure~\ref{fig1} (top) shows the TPD spectra of pure \ce{CH3OH} compared to the TPD spectra of the \ce{CH4} + \ce{H} + $^{18}$\ce{O2} reaction (bottom) for \emph{m/z\/} = 31, 32 and 33, 34, respectively. We note that \emph{m/z\/} = 32 also represents the \ce{O2}$^{.+}$ ion in the \ce{CH4} + \ce{H} + \ce{O2} reaction, and therefore TPD data from the $^{18}$\ce{O2} isotope experiment are used instead. Because \emph{m/z\/} = 31 (\ce{CH3O}$^{+}$) and 32 (\ce{CH3OH}$^{.+}$) represent the most intense ions for the \ce{CH3OH} cracking pattern found in our experimental setup and because their isotopically--induced shifted \emph{m/z\/} values in the isotope--enriched experiments can be tracked for most experiments (i.e., the \emph{m/z\/} values remain characteristic of \ce{CH3OH} and not of other species), we use the \emph{m/z\/} values representing \ce{CH3O}$^{+}$ and \ce{CH3OH}$^{.+}$ to confirm the formation of \ce{CH3OH}. As seen in Figure~\ref{fig1} (top), the signals for \emph{m/z\/} = 31 and 32 peak at 148 K for pure \ce{CH3OH}. The same \emph{m/z\/} signals peak at 150 K in the \ce{CH4} + \ce{H} + $^{18}$\ce{O2} experiment, as seen in Figure~\ref{fig1} (bottom). This slight increase in the \ce{CH3OH} desorption temperature is expected since the binding energy of \ce{CH3OH} will be influenced by the presence of \ce{H2O}. It should be noted that \ce{H2O}, along with \ce{H2O2}, are formed during the experiments and are not initial reactants of the ice mixture. The similar peak positions and profiles for the \ce{CH3O}$^{+}$ (\ce{CH3}$^{18}${O}$^{+}$) and \ce{CH3OH}$^{.+}$ (\ce{CH3}$^{18}$\ce{OH}$^{.+}$) ion signals between the two experiments allows to assign the observed desorption as originating from \ce{CH3OH}. 

\begin{figure}
\resizebox{\hsize}{!}{\includegraphics{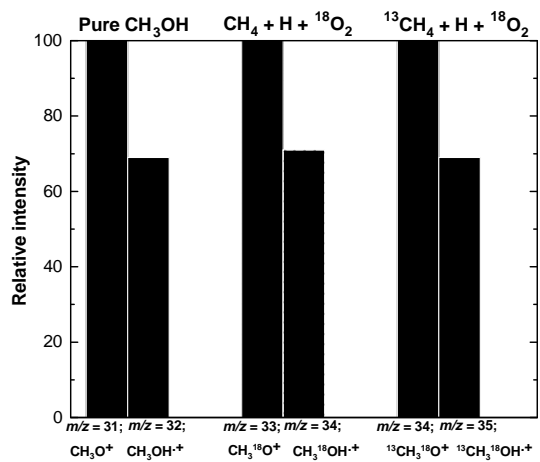}}
\caption{Integrated QMS signals normalized to the integrated QMS signals of the \ce{CH3O}$^{.+}$, \ce{CH3}$^{18}$\ce{O}$^{.+}$, and $^{13}$\ce{CH3}$^{18}$\ce{O}$^{.+}$ ions from the \ce{CH3OH} (exp. 3.0), \ce{CH4 + H} + $^{18}$\ce{O2} (exp. 2.2), and $^{13}$\ce{CH4 + H} + $^{18}$\ce{O2} (exp. 2.3) experiments, respectively. Ratios of the integrated intensities between the isotopes are in agreement with each other and with NIST values representing \ce{CH3OH}.}  
\label{fig2}
\end{figure}

This assignment is further constrained by quantitatively comparing the \ce{CH3OH} fragmentation pattern upon 70 eV electron impact ionization with values available from the NIST database\footnote{NIST Mass Spec Data Center, S.E. Stein, director, ``Mass Spectra'' in NIST Chemistry WebBook, NIST Standard Reference Database Number 69, Eds. P.J. Linstrom and W.G. Mallard, National Institute of Standards and Technology, Gaithersburg MD, 20899, doi:10.18434/T4D303, (retrieved November 7, 2017)}. Figure~\ref{fig2} presents a column chart of the relative integrated intensities of the \ce{CH3O}$^{+}$ and \ce{CH3OH}$^{.+}$ ion signals observed in a pure \ce{CH3OH} experiment and in isotopically--enriched $^{13}$\ce{CH4 + H} + $^{18}$\ce{O2} experiments. As seen in Figure~\ref{fig2}, the relative integrated intensities of pure \ce{CH3OH} for \emph{m/z\/} = 31 (\ce{CH3O}$^{+}$) and 32 (\ce{CH3OH}$^{.+}$) are 100 and 69, respectively. In the \ce{CH4 + H} + $^{18}$\ce{O2} experiment, the intensities for \emph{m/z\/} = 33 (\ce{CH3}$^{18}$\ce{O}$^{+}$) and 34 (\ce{CH3}$^{18}$\ce{OH}$^{.+}$) are 100 and 71, respectively. In the $^{13}$\ce{CH4 + H} + $^{18}$\ce{O2} experiment, the values for \emph{m/z\/} = 34 ($^{13}$\ce{CH3}$^{18}$\ce{O}$^{+}$) and 35 ($^{13}$\ce{CH3}$^{18}$\ce{OH}$^{.+}$) are 100 and 69, respectively. In the NIST database, the relative integrated intensities for \emph{m/z\/} = 31 (\ce{CH3O}$^{+}$) and 32 (\ce{CH3OH}$^{.+}$) are 100 and 74, respectively. The relative integrated intensities in the three experiments are nearly identical and are close to the values from NIST, fully in line with \ce{CH3OH} formation in the ice. 

\subsubsection{RAIRS}
\label{3.1.2.}

\begin{figure}
\resizebox{\hsize}{!}{\includegraphics{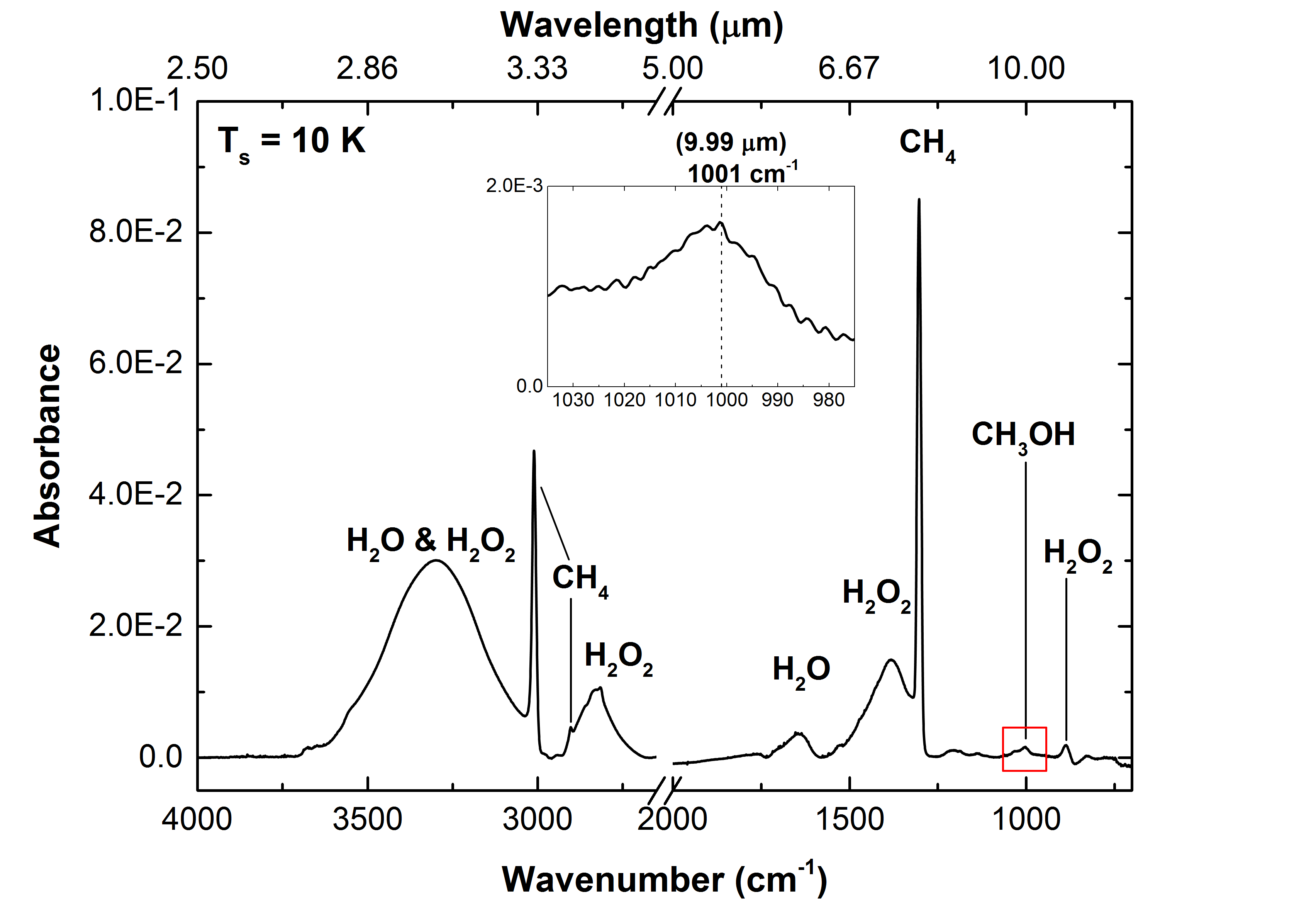}}
\caption{RAIR spectrum of the reaction, \ce{CH4 + H + O2}, on a 10 K substrate surface after 12 hours (exp. 1.0). The relative column density of \ce{CH4}:\ce{CH3OH} is 100:1, and the relative column density of \ce{H2O}:\ce{CH3OH} is 100:4. Spectra of the resulting species are shown (see Table~\ref{table2}), most noticeably a weaker band around 1000 cm$^{-1}$ (inset) highlighting the \ce{C-O} stretch of the \ce{CH3OH} feature.}
\label{fig3}
\end{figure}

Figure~\ref{fig3} shows the resulting 4000--700 cm$^{-1}$ (2.5--14.3 $\mu$m) spectrum for \ce{CH4 + H + O2} interacting on a 10 K substrate. Table~\ref{table2} lists the frequencies measured, as well as the identification of species observed and the corresponding vibrational modes. The assignment of the peaks are constrained by considering literature values, isotope experiments, ice desorption temperatures, and varying flux ratios of the reactants. The inset in Figure~\ref{fig3} shows a band at 1001 cm$^{-1}$ that is assigned to the \ce{C-O} stretching mode of \ce{CH3OH}. This assignment is based on experiments performed in Section~\ref{spectral}. Literature values for this band \citep{wada2006methanol,hodyss2009photochemistry,martin2016study} are normally somewhat higher; around 1015 cm$^{-1}$ when \ce{CH3OH} is mixed with \ce{H2O}. Even though this band is rather weak, it offers a tool to unambiguously link to \ce{CH3OH} as it has no overlap with bands of other species formed in these experiments and actually provides a nicely isolated feature.

\begin{figure}
\resizebox{\hsize}{!}{\includegraphics{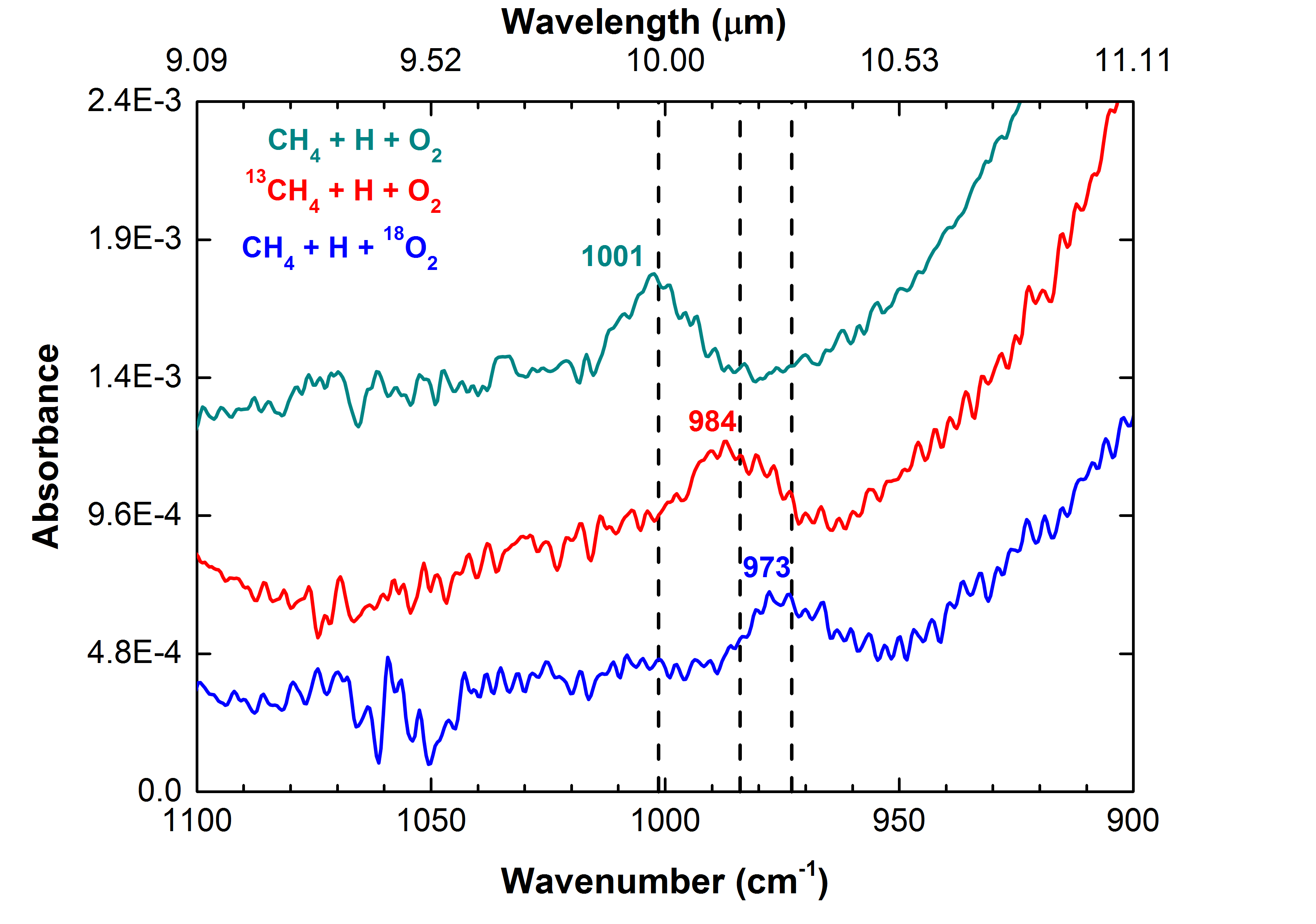}}
\caption{RAIR spectra of the reactions, \ce{CH4 + H + O2} (exp. 2.0), $^{13}$\ce{CH4} + H + \ce{O2} (exp. 2.1), and \ce{CH4 + H} + $^{18}$\ce{O2} (exp. 2.2) on a 10 K substrate surface after 6 hours each.} 
\label{fig4}
\end{figure}

A series of additional experiments are performed in order to further prove that the band at 1001 cm$^{-1}$ is indeed due to the \ce{C-O} stretching frequency of \ce{CH3OH}. This is realized in a set of isotope substitution experiments. Figure~\ref{fig4} shows spectra of three experiments that involve \ce{CH4}, $^{13}$\ce{CH4}, \ce{O2}, and $^{18}$\ce{O2} isotopes. By the simple harmonic oscillator approach, a heavier isotope should result in a redshift in the stretching frequency. As seen in Figure~\ref{fig4}, the 1001 cm$^{-1}$ feature redshifts to 984 cm$^{-1}$ and 973 cm$^{-1}$ for the reactions involving $^{13}$\ce{CH4} and $^{18}$\ce{O2}, respectively. This leads to differences of 17 cm$^{-1}$ and 28 cm$^{-1}$ from the 1001 cm$^{-1}$ feature, respectively, comparable to the differences found in \ce{CH3OH} isotope experiments performed by \citet{maity2014infrared}. These findings are consistent with isotopically enriched \ce{C-O} bonds in newly formed \ce{CH3OH}.

   \begin{table*}
\caption{Infrared ice signatures from the \ce{CH4 + H + O2} reaction and the corresponding vibrational modes.} 
\label{table2} 
\centering 
\begin{tabular}{c c c c c} 
\hline\hline 
Peak position & Peak position & Molecule & Mode$^{*}$ & Reference\\
(cm$^{-1}$) & ($\mu$m) \\ 
\hline 
884 & 11.3 & \ce{H2O2} & $\upsilon_3$ & 1, 2, 3 \\ 
1001 & 9.99 & \ce{CH3OH} & $\upsilon_8$ & this work \\
1302 & 7.68 & \ce{CH4} & $\upsilon_4$ & 4, 5, 7, 8, 9 \\
1381 & 7.24 & \ce{H2O2} & $\upsilon_2$ & 1, 2, 3 \\
1637 & 6.11 & \ce{H2O} & $\upsilon_2$ & 6 \\
2815 & 3.55 & \ce{CH4} & $\upsilon_2$ + $\upsilon_4$ & 4, 5, 7, 8, 9\\ 
2836 & 3.53 & \ce{H2O2} & 2$\upsilon_2$ & 1, 2, 3\\
2902 & 3.45 & \ce{CH4} & $\upsilon_1$ & 7, 8, 9\\
3010 & 3.32 & \ce{CH4} & $\upsilon_3$ & 4, 5, 7, 8, 9\\
3295 & 3.0 & \ce{H2O2} + \ce{H2O} & \ce{O-H} stretch & 10, 11\\
3675 & 2.7 & \ce{H2O} & dangling bonds & 7\\
\hline 
\end{tabular}

\tablefoot{
\tablefoottext{*}{The vibrational mode numbers are obtained from NIST.}
}

\tablebib{
(1) \citet{giguere1959infrared}; (2) \citet{lannon1971infrared}; (3) \citet{romanzin2011water}; (4) \citet{hagen1981infrared}; (5) \citet{gerakines2005strengths}; (6) \cite{hodyss2009photochemistry}; (7) \citet{galvez2009spectroscopic}; (8) \citet{herrero2010interaction}; (9) \citet{ennis2011chemical}; (10) \citet{cuppen2010water}; (11) \citet{ioppolo2010water}.}
\end{table*}

\subsubsection{Temperature--dependent RAIR difference spectra}
\label{3.1.3.}

\begin{figure}
\resizebox{\hsize}{!}{\includegraphics{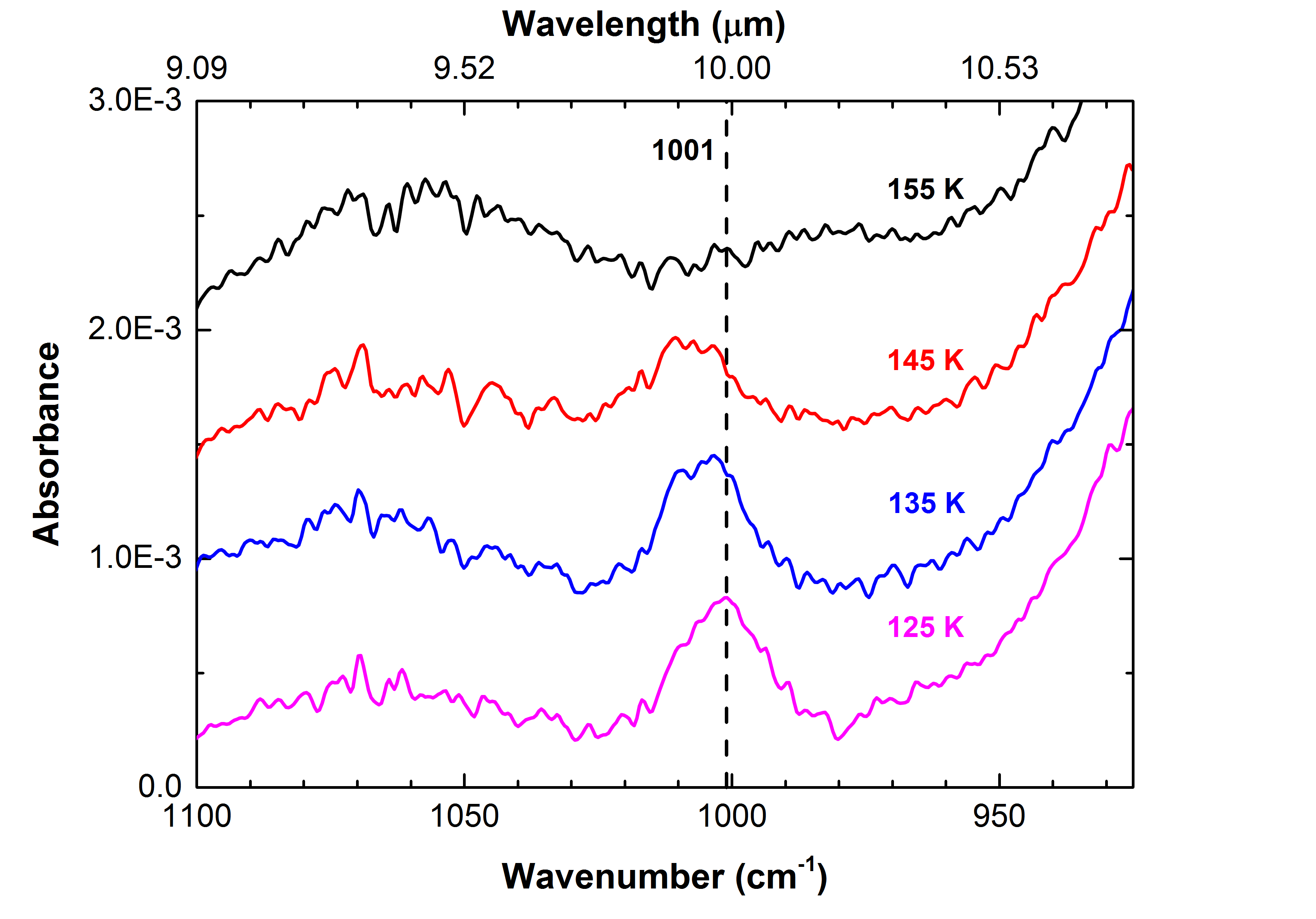}}
\caption{RAIR difference spectra of the \ce{CH4 + H + O2} reaction (exp. 2.0). The 1001 cm$^{-1}$ feature blueshifts between 125 K and 145 K, and could be due to \ce{CH3OH} segregating in a \ce{H2O}--ice environment \citep{martin2014thermal}, and disappears upon further heating to 155 K.}
\label{fig5}
\end{figure}

The TPD and RAIRS data can be correlated to one another to match IR features with desorption temperatures to further identify species that are initially made in the ice (i.e., before thermal processing via the TPD technique). Figure~\ref{fig5} shows the change in the IR features as a function of the substrate temperature (i.e., RAIR difference spectra) for the \ce{CH4 + H + O2} reaction. In Section~\ref{3.1.1.} the desorption of \ce{CH3OH} is observed at 150 K, and in Section~\ref{3.1.2.} the IR feature at 1001 cm$^{-1}$ is shown to be due to \ce{CH3OH} formation. The combination of these two pieces of data tells us that below 150 K, \ce{CH3OH} should be present in the ice, and above 150 K a majority of the \ce{CH3OH} should desorb. In Figure~\ref{fig5}, the \ce{C-O} bond feature remains present at a substrate temperature of 145 K and is absent at a substrate temperature of 155 K. This correlates well with the desorption temperature of \ce{CH3OH} observed in Figure~\ref{fig1} (bottom). Thus, the experiments discussed in Sections~\ref{3.1.1.}, ~\ref{3.1.2.}, and ~\ref{3.1.3.} make a consistent case that \ce{CH3OH} is formed in the ice.

\begin{figure}
\resizebox{\hsize}{!}{\includegraphics{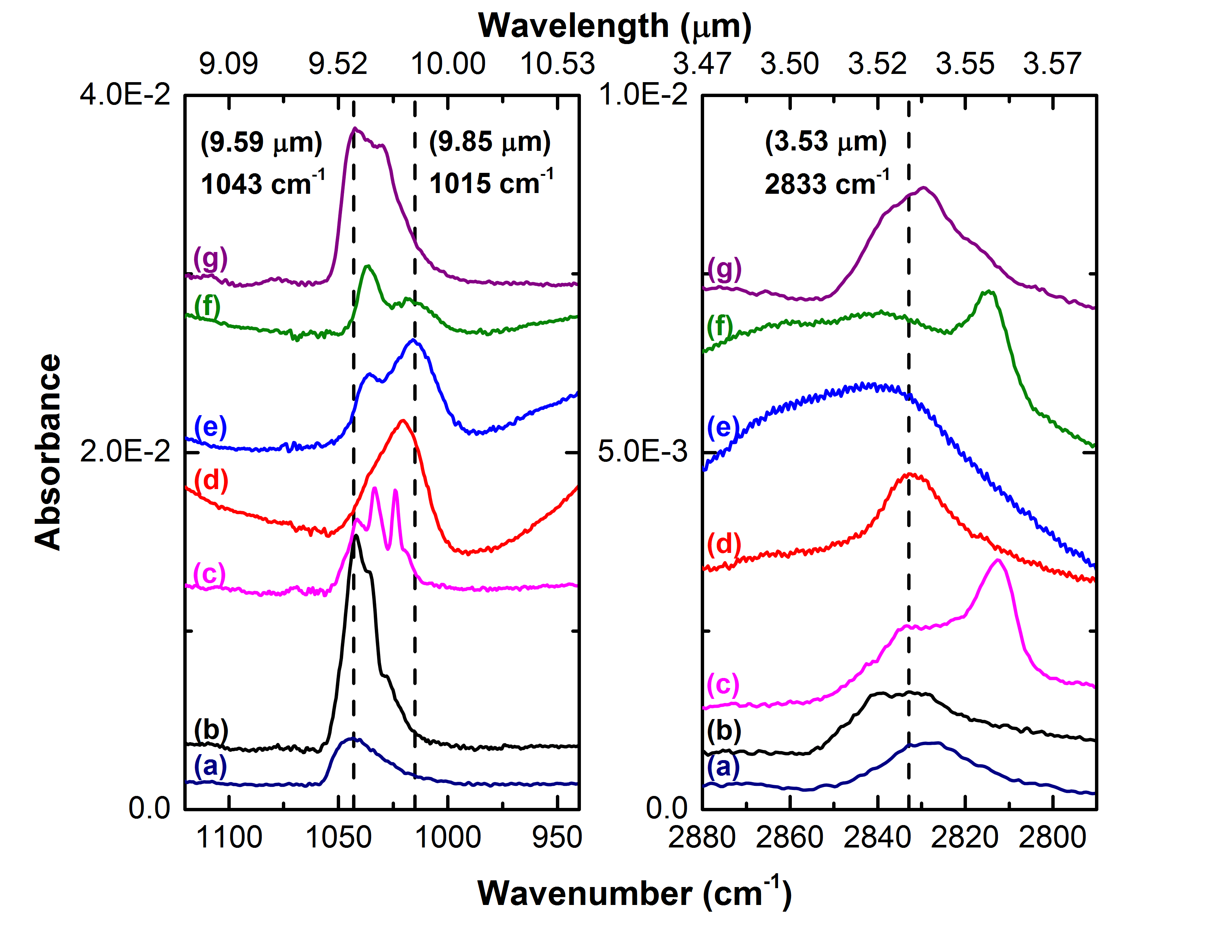}}
\caption{(Left) RAIR spectra of the \ce{C-O} stretch of \ce{CH3OH} when \ce{CH3OH} is mixed with different species. (a) Pure \ce{CH3OH} (exp. 4.0), (b) \ce{CH3OH} + \ce{O2} (exp. 4.1), (c) \ce{CH3OH} + \ce{CH4} (exp. 4.2), (d) \ce{CH3OH} + \ce{H2O} (exp. 4.3), (e) \ce{CH3OH} + \ce{H2O} (exp 4.4), (f) \ce{CH3OH} + \ce{H2O} + \ce{CH4} (exp 4.5), and (g) \ce{CH3OH} + \ce{CO} (exp 4.6). (Right) RAIR spectra of the symmetric \ce{C-H} stretch of \ce{CH3OH}.}
\label{fig6}
\end{figure}

\subsection{Spectral signature of \ce{CH3OH} in a \ce{H2O}--rich interstellar ice analogue}
\label{spectral}

Because the \ce{C-O} bond of \ce{CH3OH} is known to be sensitive to its surrounding environment, it holds a potential as a diagnostic tool to provide a deeper insight to the interactions of \ce{CH3OH} with other species in the \ce{H2O}--rich interstellar ice analogue. Figure~\ref{fig6} (left) shows a RAIR spectrum of the \ce{C-O} stretching frequency of pure \ce{CH3OH}  (a) in comparison to \ce{CH3OH} embedded in an environment originating from the \ce{CH4 + H + O2} reaction  (f), in addition to \ce{CH3OH} mixed with other species (b--e, g). Figure~\ref{fig6} (right) shows a RAIR spectrum of the \ce{C-H} stretch of pure \ce{CH3OH}; a mode that is a useful tracer of \ce{CH3OH} ice and can be probed by observational facilities (e.g., JWST NIRSpec). The experiments are performed using both the previously described HABS and a microwave atom source (MWAS), which dissociates a fraction of the incoming molecules into fragments. The fragments can recombine with each other and form products in the ice that would be found in, for example, the \ce{CH4 + H + O2} reaction (\ce{OH}, H, \ce{O2}, \ce{O}, \ce{H2}, \ce{H2O2}, etc.). The \ce{C-O} stretching frequency of pure \ce{CH3OH} is found at ~1043 cm$^{-1}$ and redshifts to ~1015 cm$^{-1}$ in the ice where \ce{CH4} and fragments from dissociated \ce{H2O} are co--deposited. We note that the peak does not redshift to 1001 cm$^{-1}$, and there is a logical explanation for this. As illustrated in Figure 7 in the study by ~\citet{dawes2016using}, an increase in the \ce{H2O} concentration in a \ce{CH3OH + H2O} mixture does not necessarily correlate to a more redshifted \ce{C-O} stretching frequency. Thus, it is expected that a specific ratio of the ice products in the \ce{CH4 + H + O2} reaction is needed in order to recreate the 1001 cm$^{-1}$ feature. From the spectra presented in Figure~\ref{fig6}, it is clear that the largest redshift for the \ce{C-O} stretch takes place in a \ce{H2O}--rich ice matrix that contains \ce{CH4}, \ce{H2O}, and \ce{H2O} dissociation products (i.e., the mixture in the actual \ce{CH4 + H + O2} experiment; exp. 4.5), as expected. When \ce{CH3OH} is mixed with a single species, such as \ce{H2O} or \ce{CH4}, the \ce{C-O} stretching frequency does not redshift as much. In conclusion, it is proposed that many, rather than one or two molecular species, interact simultaneously with \ce{CH3OH} in the \ce{CH4 + H + O2} reaction. 

\subsection{\ce{CH3OH} formation at 10 and 20 K}
\label{3.2}

As mentioned in Section ~\ref{2.}, an absolute value for the \ce{CH3OH} column density is not obtained in this study. Thus, in order to provide information on the formation yield of \ce{CH3OH} formed at 10--20 K, a relative yield is obtained by comparing the amount of \ce{CH3OH} formed at 10 K (exp. 2.0) to the amount of \ce{CH3OH} formed at 20 K (exp. 5.0). Since it is expected in interstellar space that \ce{CH4} and \ce{H2O} ice are initially formed by hydrogenation reactions \citep{tielens1982model,oberg2008c2d,ioppolo2008laboratory,cuppen2010water,van2013interstellar}, the \ce{CH4}:\ce{H2O} ice ratio should not be drastically different between surface temperatures of 10 and 20 K. Therefore, the \ce{CH4}:\ce{H2O} ice ratio is kept the same in this temperature range. At 20 K, a 65\% decrease in the \ce{CH3OH} abundance is observed compared to the amount of \ce{CH3OH} formed at 10 K. This decrease in abundance is expected to be due to the decrease in the H--atom lifetime on the surface \citep{fuchs2009hydrogenation,cuppen2010water}, as expected if \ce{CH3OH} is being formed by the \ce{OH} radical. Although, it should be noted that the reaction steps for \ce{CH3OH} formation may also be affected by the temperature change, which can influence the change in the \ce{CH3OH} abundance. 

\subsection{Constraining the formation of \ce{CH3OH} in the overall reaction network}
\label{3.3}

\begin{figure}
\resizebox{\hsize}{!}{\includegraphics{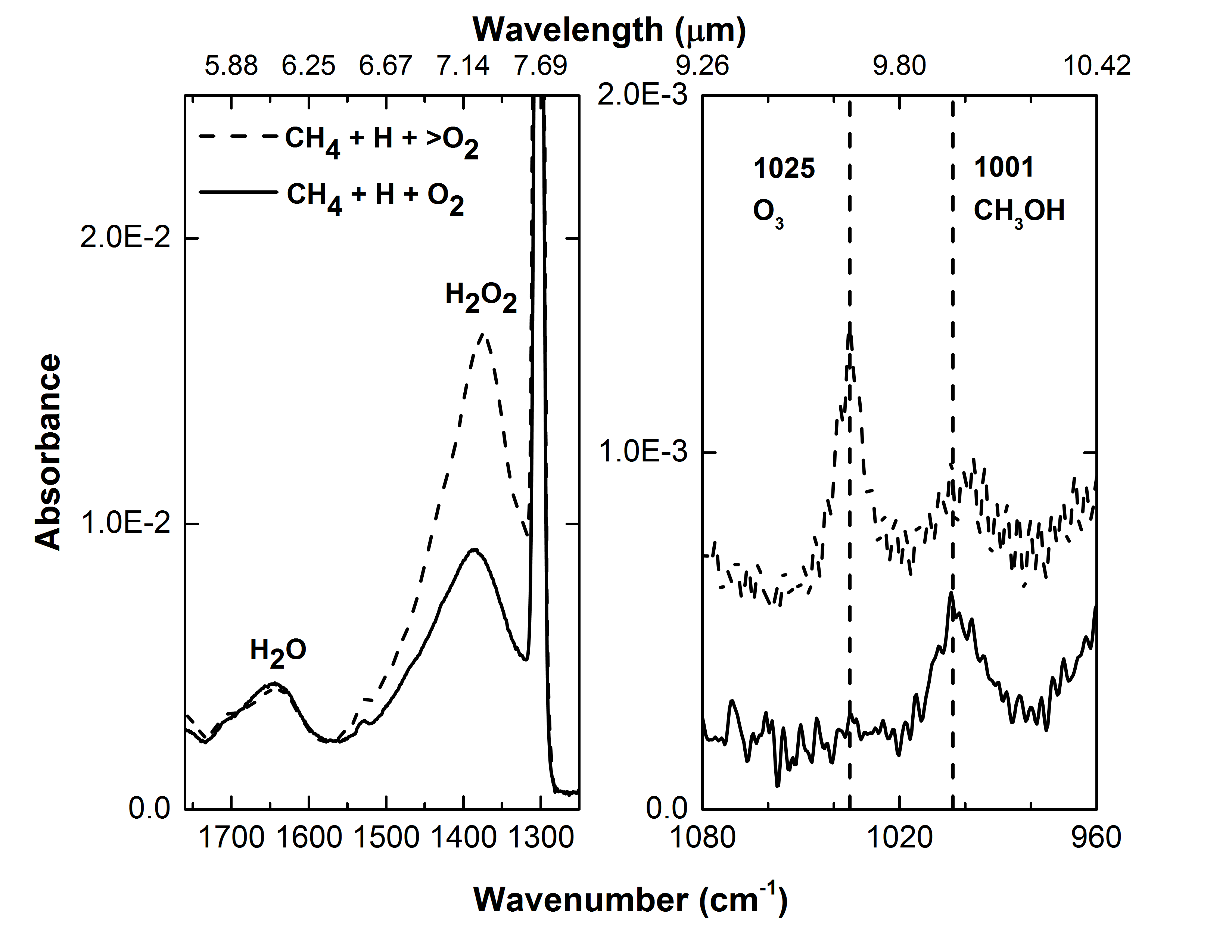}}
\caption{RAIR spectra of the reactions, \ce{CH4 + H + O2}, where >\ce{O2} indicates a threefold increase in the \ce{O2} flux. The dashed--line spectrum corresponds to exp. 2.0, and the solid--line spectrum corresponds to exp. 6.0.}
\label{fig7}
\end{figure}

The experiments discussed here show that it is possible to form \ce{CH3OH} in a \ce{H2O}--rich environment along the low temperature pathway:
\begin{subequations}\label{eq1}
\begin{equation}
\ce{CH4 + OH} \rightarrow \ce{CH3 + H2O} 
\end{equation}
\begin{equation}
\ce{CH3 + OH} \rightarrow \ce{CH3OH}
\end{equation}
\end{subequations}
To observe how these channels fit into the overall reaction network ongoing in \ce{H2O}--rich ice, the change in abundances of \ce{CH3OH} and other ice species needs to be tracked \citep{oberg2010effect}. Figure~\ref{fig7} shows two RAIR spectra of the reaction, \ce{CH4 + H + O2}, and the subsequent infrared features pertaining to newly formed \ce{CH3OH}, \ce{H2O}, \ce{H2O2}, and \ce{O3}. With an enhancement of the \ce{O2} flux (dashed line  in Figure~\ref{fig7}), the \ce{CH3OH} abundance decreases and the \ce{H2O2} and \ce{O3} abundances clearly increase. The decrease in the \ce{CH3OH} abundance at the expense of the \ce{H2O2} and \ce{O3} abundances gives an indication that the formation channel of \ce{CH3OH} at least crosses with the formation channels of \ce{H2O2} and \ce{O3}.

The link between the formation of \ce{CH3OH}, \ce{H2O2}, and \ce{O3} in the \ce{H2O}--rich ice can be better understood when looking at the reaction network presented by \citet{cuppen2010water}. In this study, the \ce{H + O2} reaction is extensively discussed, and it is found that the \ce{O} atom used to react with \ce{O2} to form \ce{O3} is most likely created by two reactions, 
\begin{equation}\label{eq2}
\ce{H + HO2} \rightarrow \ce{H2O + O}
\end{equation}
and
\begin{subequations}\label{eq3}
\begin{equation}\label{eq3a}
\ce{OH + O2} \rightarrow \ce{HO2 + O}
,\end{equation}
where \ce{OH} is mainly formed by the following reaction:
\begin{equation}\label{eq3b}
\ce{H + HO2} \rightarrow \ce{2OH}
\end{equation}
\end{subequations}

\begin{figure}
\resizebox{\hsize}{!}{\includegraphics[scale=1.5]{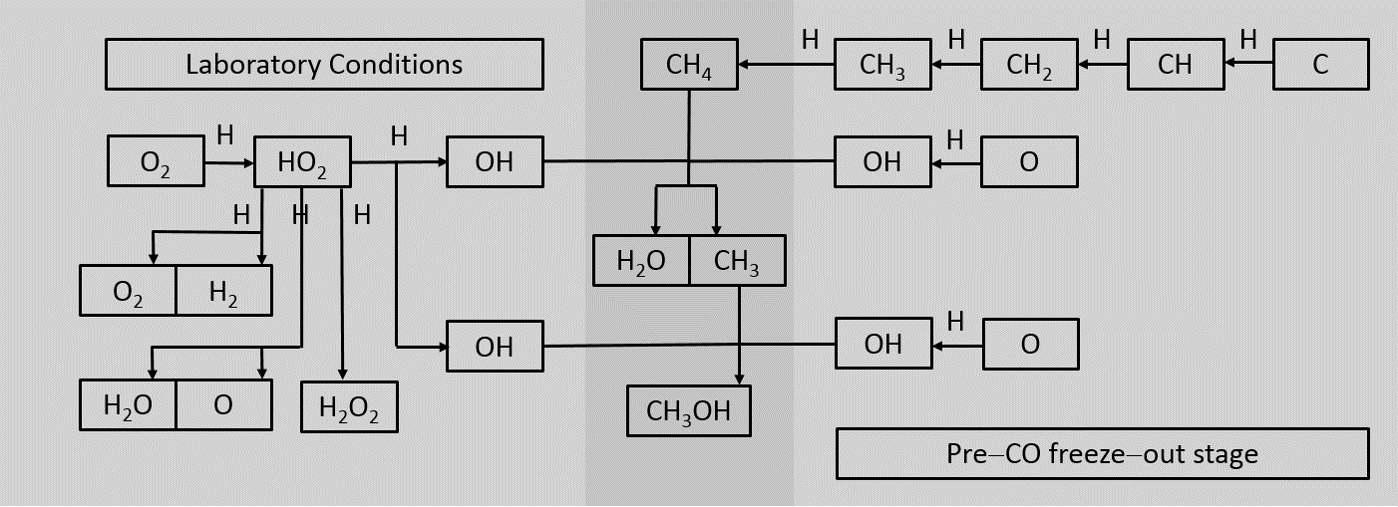}}
\caption{Comparison of the laboratory work discussed here and the astrochemical network, resulting in the formation of \ce{CH3OH} in the pre--CO freeze--out stage. Only the relevant steps are shown.}
\label{fig8}
\end{figure}

We note that only the relevant steps are discussed here, and the full reaction scheme can be found in \citet{cuppen2010water}. The RAIR spectra shown in Figure~\ref{fig7} suggest that the latter reactions are dominant in the \ce{CH4 + H + O2} reaction. As observed in Figure~\ref{fig7}, a threefold increase in the \ce{O2} flux (exp. 2.0 has a threefold increase in the \ce{O2} flux compared to exp. 6.0) results in a substantial increase in the \ce{H2O2} abundance compared to the change in the \ce{H2O} abundance. This is consistent with reactions~\ref{eq3a} and~\ref{eq3b} as \ce{HO2} is an intermediate of the \ce{H2O2} product; \ce{HO2} is also an intermediate of \ce{H2O} formation, as shown in reaction~\ref{eq2}. However, comparison of the two spectra in Figure~\ref{fig7} shows that an increase in the \ce{O3} abundance is followed by a significant difference in the \ce{H2O2} abundance and a relatively minor change in the \ce{H2O} abundance, suggesting that reactions~\ref{eq3a} and~\ref{eq3b} are carried out more efficiently than reaction~\ref{eq2} in the \ce{CH4 + H + O2} experiment. Decreasing the \ce{CH4} abundance by 20\% shows a 10\% decrease in the \ce{CH3OH} abundance and no change in the \ce{H2O} abundance, which further supports that reaction~\ref{eq2} is not the dominating channel. In addition, reactions~\ref{eq3a} and~\ref{eq3b} show that \ce{O3} is formed by an \ce{OH}--induced process, and this is expected since \ce{CH3OH} is formed by an \ce{OH}--mediated reaction. These insights piece together to demonstrate that reactions~\ref{eq3a} and~\ref{eq3b} are the likely pathways that connect the formation channel of \ce{CH3OH} to other formation channels in a \ce{H2O}--rich ice reaction network. In Figure~\ref{fig8}, the laboratory experiments and the astrochemical proposed reactions are summarized in one figure. The main difference between the two is how \ce{OH} radicals are inserted into the reaction scheme, but this does not change the conclusions derived here for the presented pre--CO freeze--out formation scheme.   

\section{Astrophysical implications}
\begin{figure}
\resizebox{\hsize}{!}{\includegraphics{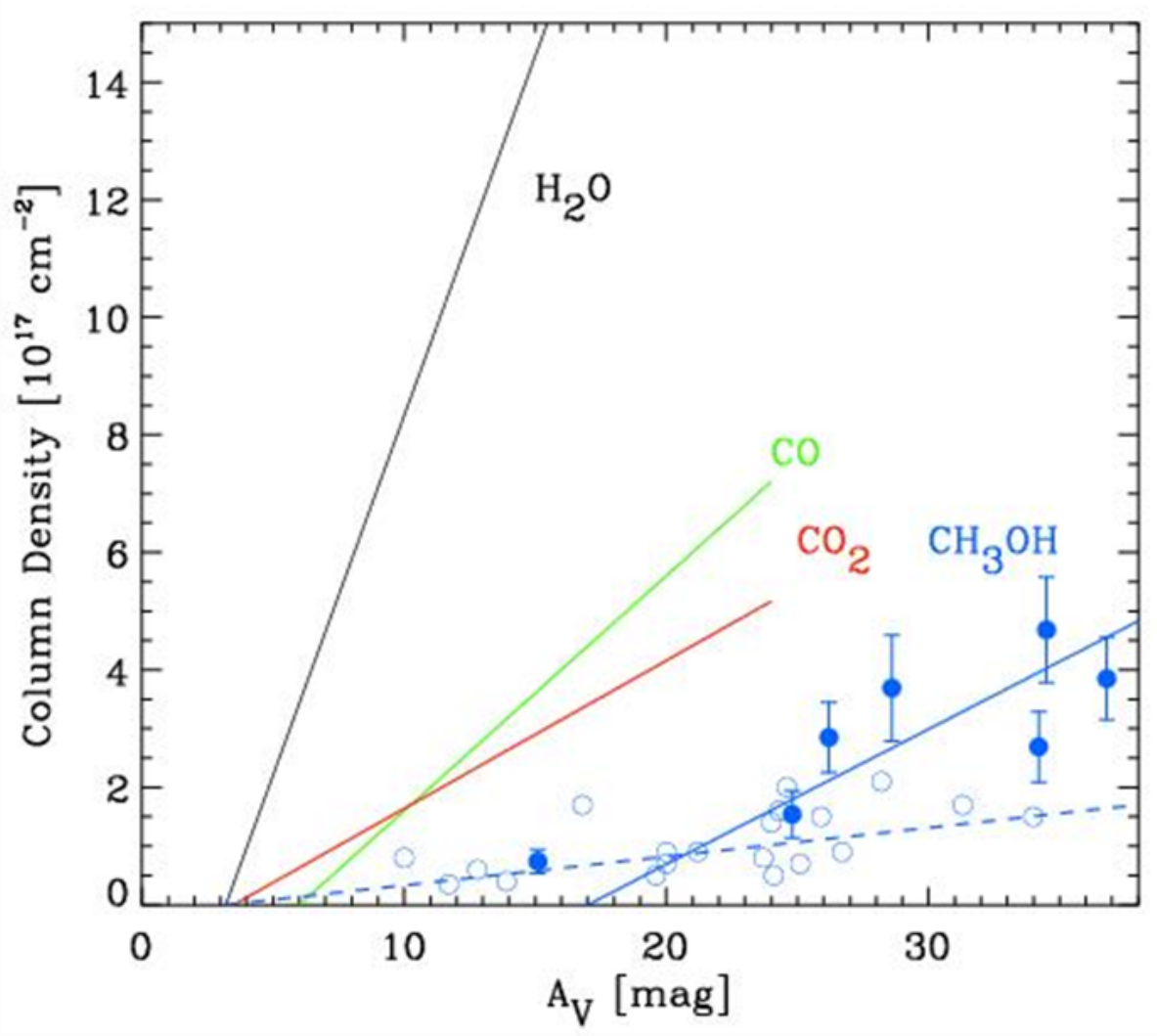}}
\caption{Relationships between ice column density and visual extinction ($A_V$) as observed toward stars behind dense clouds and cores for \ce{H2O} (black), \ce{CO} (green), \ce{CO2} (red), and \ce{CH3OH} (blue). For clarity, for \ce{H2O}, \ce{CO}, and \ce{CO2}, the least--squares linear fits are shown and not the actual data points. For \ce{CH3OH}, blue bullets indicate detections with 1--$\sigma$ error bars and open circles indicate 3--$\sigma$ upper limits. The solid blue line is a fit to the detections, and the dashed blue line (from exp. 1.0) indicates the scenario in which \ce{CH3OH} has the same ice formation threshold as \ce{H2O} at a maximum abundance of \ce{CH3OH}:\ce{H2O} = 4\%, as found in the listed experiments. The 3--$\sigma$ upper limits prove in principle that \ce{CH3OH}:\ce{H2O} is less than 4\% in the observed sight lines. This figure was adapted from Figure 7 in \citet{boogert2015observations}.}
\label{fig9}
\end{figure}

The experimental work presented here calls into question the $A_V$ at which \ce{CH3OH} ice starts to grow (i.e., the \ce{CH3OH} ice formation threshold). If an observational search for the \ce{CH3OH} ice formation threshold in low extinction (i.e., \ce{H2O}--rich) environments is to be performed, it would be useful to know beforehand an approximation of the \ce{CH3OH} abundance in these \ce{H2O}--rich ices. That way, the amount of \ce{CH3OH} formed before and after the \ce{CO} freeze--out stage can be distinguished. A laboratory experiment of \ce{CO + H} performed under similar conditions to exp. 2.0 shows that the \ce{CO}:\ce{CH3OH} ratio is 100:20. In the parallel \ce{CH4 + OH} experiment, the \ce{CH4}:\ce{CH3OH} ratio is 100:1. A comparison of these two experiments leads to the conclusion that the \ce{CH4 + OH} reaction is 20 times less efficient at producing \ce{CH3OH} than the \ce{CO + H} reaction in the laboratory setting. Making the simple assumption that \ce{CH3OH} is primarily formed by \ce{CH4 + OH} and \ce{CO + H} in interstellar ices around the \ce{CO} freeze--out stage, laboratory data can be combined with observational data to determine the amount of \ce{CH3OH} in the two different ice phases. According to \citet{boogert2015observations}, the \ce{H2O}:\ce{CH4} and \ce{H2O}:\ce{CO} median ice ratios around low mass young stellar objects (LYSOs) are 100:4.5 and 100:21, respectively. Incorporation of the laboratory results shows that 0.045\% of the interstellar ice should contain \ce{CH3OH} that derives from \ce{CH4}, and 4.2\% of the ice should contain \ce{CH3OH} that derives from \ce{CO}. Thus, of the total observed abundance of \ce{CH3OH}, about 1\% of the total amount is the abundance of \ce{CH3OH} from the \ce{CH4 + OH} reaction that is expected to be ongoing in the \ce{H2O}--rich ice phase. Although a \ce{CH3OH}:\ce{H2O} percentage of 0.045\% may appear insignificant, multiplying this value by the typical \ce{H2O} ice to \ce{H} ratio of 2E--5 yields a \ce{CH3OH} ice to \ce{H} ratio of 9E--9, which is greater than the abundance of gas--phase \ce{CH3OH} and other gas--phase species detected in cold cores relative to \ce{H2} \citep{herbst2009complex}. Using values from the 20 K experiment (exp. 5.0), the \ce{CH3OH} ice to \ce{H} ratio results in a value of 3E--9, which is similar to the abundance of gas--phase \ce{CH3OH} in these particular astrophysical environments. We note that the median ice ratios contain sources that also have a relatively high amount of \ce{CO} ice, so the calculated values are not entirely representative of low extinction (i.e., \ce{H2O}--rich ice) scenarios. See Figure~\ref{fig9} for a better approximation of the possible \ce{CH3OH} abundance in the \ce{H2O}--rich ice layer at various extinctions.  

Figure~\ref{fig9}, which is adapted from \citet{boogert2015observations}, shows that \ce{CH3OH} detections have an ice formation threshold at an $A_V$ value higher than \ce{H2O} and \ce{CO} (the ice formation threshold in this figure is the x--axis cut--off divided by 2 because the background stars trace both the front and back sides of the clouds). This is in line with the currently accepted model that \ce{CH3OH} ice is mainly formed in prestellar cores at $A_V$ > 9 via hydrogenation of accreted \ce{CO} molecules. However, what is puzzling is that the upper limit sight lines are all below the detections and at similar $A_V$ values to that of the detections. The \ce{CH3OH}:\ce{H2O} ratio of 4:100 derived from the laboratory work presented here (from exp. 1.0) is represented by the dashed blue line and is at the level of 3---$\sigma$ upper limits. Thus, the current observations are consistent with a  \ce{CH3OH}:\ce{H2O} ratio of at most  a few percent, though likely not as high as 4\%, the value seen in our optimized experiments, since \ce{H2O} is a by--product in the \ce{CH4 + H + O2} reaction and is not the main source of \ce{H2O} in interstellar ices. At this level, a formation threshold as low as that of \ce{H2O} (1.6 mag) as a result of chemical reactions in \ce{H2O}--rich ices cannot be excluded. More sensitive observations are needed to further constrain this scenario and with the upcoming launch of the JWST, this soon will be within range.

The insights gained here also link to the formation of larger COMs. Reaction~\ref{eq1} gives rise to the possibility that such a mechanism could lead to the formation of more complex species in \ce{H2O}--rich ices. It has been shown that COMs such as glycolaldehyde, ethylene glycol, and glycerol can be successfully formed by hydrogenation of \ce{CO} \citep{fedoseev2015experimental,fedoseev2017formation}, thus providing a promising reaction scheme for COM formation in \ce{CO}--rich interstellar ices. Formation of COMs  in \ce{H2O}--rich ices has been less studied, and the mechanism described in this paper could be a potential source of molecular complexity in cold \ce{H2O}--dominated ice matrices, although formation efficiencies will be lower.       
\label{4.}

\section{Conclusions}
\label{5.}

The formation of \ce{CH3OH} by \ce{CH4 + OH} shown in this laboratory study suggests that \ce{CH3OH} ice can also be formed before the heavy \ce{CO} freeze--out stage in prestellar cores, i.e., at $A_V$ < 3 and grain temperatures of 20 K. The main findings from this work are summarized below:

\renewcommand\labelitemi{\ding{117}}
\begin{itemize}
\setlength\itemsep{1em}
 
\item The formation of \ce{CH3OH} in \ce{H2O}--rich ices occurs by the sequential reactions, \ce{CH4 + OH}$\,\to\,$\ce{CH3 + H2O} and \ce{CH3 + OH}$\,\to\,$\ce{CH3OH}.

\item Since \ce{CH4} and \ce{OH} radicals are expected to be found in the \ce{H2O}--rich ice phase of cold dense cores, \ce{CH3OH} is also expected to be found in those ices, suggesting that the \ce{CH3OH} formation threshold is below ~$A_V$ = 9.  

\item Raising the sample temperature to 20 K results in a 65\% decrease in the \ce{CH3OH} abundance compared to \ce{CH3OH} formed on a 10 K surface, showing that the formation pathway of \ce{CH3OH} in this study is relevant to the period of the cold dense core stage at which ice species are formed primarily by atom--induced reactions (i.e., before gas--phase molecules, like CO, accrete onto the grain surface).

\item Under similar experimental parameters, the \ce{CO + H} channel is about 20 times more efficient to forming \ce{CH3OH} than the \ce{CH4 + OH} channel at temperatures around 10 K.

\item More sensitive astronomical observations are warranted to determine the \ce{CH3OH}:\ce{H2O} ratio in \ce{H2O}--rich interstellar ices, which is likely to be a few percent at most, as derived from Figure~\ref{fig9}. 

\item The formation of COMs under cold dark cloud conditions is mainly linked to the \ce{CO} freeze--out stage. The reactions studied here carry the potential to lead to the formation of COMs at an earlier astrochemical evolution stage. 
\end{itemize}
\begin{acknowledgements}
This research was funded by the Dutch Astrochemistry Network II (DANII), through a VICI grant of NWO (the Netherlands Organization for Scientific Research), and A--ERC grant 291141 CHEMPLAN. The financial support by NOVA (the Netherlands
Research School for Astronomy) and the Royal Netherlands Academy of Arts and Sciences (KNAW) through a professor prize is acknowledged. G.F. acknowledges the financial support from the European Union's Horizon 2020 research and innovation programme under the Marie Sklodowska--Curie grant agreement n. 664931. S.I. acknowledges the Royal Society for financial support and the Holland Research School for Molecular Chemistry (HRSMC) for a travel grant. D.Q. is thankful to Daniel Harsono, Ewine van Dishoeck, Thanja Lamberts, and Vianney Taquet for enlightening discussions leading to this publication.   

\end{acknowledgements}
\bibliography{CH3OH}
\end{document}